\documentclass[fleqn,usenatbib,useAMS]{mnras}

\usepackage{graphicx}	
\usepackage{amsmath}	
\usepackage{multicol}        
\usepackage{bm}		
\usepackage{pdflscape}	

\usepackage[switch]{lineno} 

\usepackage{newtxtext,newtxmath}

\usepackage[T1]{fontenc}

\DeclareRobustCommand{\VAN}[3]{#2}
\let\VANthebibliography\thebibliography
\def\thebibliography{\DeclareRobustCommand{\VAN}[3]{##3}\VANthebibliography}

\usepackage{float}

\makeatletter
\def\fps@figure{htbp}
\def\figure{\@ifnextchar[{\@figurewithoptions}{\@figurewithoutoptions}}
\def\@figurewithoptions[#1]{\def\baselinestretch{1}\let\@makecaption=\SFB@makefigurecaption\@float{figure}[#1]\fstyle@figure}
\def\@figurewithoutoptions{\def\baselinestretch{1}\let\@makecaption=\SFB@makefigurecaption\@float{figure}\fstyle@figure}
\makeatother

\newcommand{\comm}[1]{#1}

\newcommand{\rsun}{$r_\odot$}

\title[Turbulence Evolution in Heliospheric Imaging]{Solar Wind Turbulence Spectra and Correlation Scale from 4 to 50 $r_\odot$: STEREO and PUNCH Observations}
\author[F. Pecora et al.]{
    F. Pecora,$^{1,2}$\thanks{E-mail: francesco.pecora@inaf.it}
    G. Nistic\`o,$^{3}$
    R. Chhiber,$^{1,6}$
    Y. Yang,$^{1}$
    S. Adhikari,$^{1}$
    J. Wang,$^{1}$
    R. Patel,$^{4}$ 
    Z. Gao,$^{1}$
\newauthor
    C. DeForest,$^{4}$
    S. E. Gibson,$^{5}$
    N. Viall,$^{6}$
    M. N. Beasley,$^{4}$ 
    R. C. Colaninno,$^{7}$ 
    J. M. Hughes,$^{4}$ 
\newauthor
    R. Killough,$^{8}$ 
    D. A. Lamb,$^{4}$ 
    G. T. Laurent,$^{4}$ 
    C. Lowder,$^{4}$ 
    J. Redfern,$^{4}$ 
    D. B. Seaton,$^{4}$ 
\newauthor
    S. Van Kooten,$^{4}$ 
    N. R. Waltham,$^{9}$ 
    W. L. Wells$^{8}$ 
    and W. H. Matthaeus$^{1}$
    \\ 
    $^{1}$Department of Physics and Astronomy, University of Delaware, Newark, DE, USA\\
    $^{2}$INAF -- Institute for Space Astrophysics and Planetology, Rome, Italy;
    $^{3}$Dipartimento di Fisica, Universit\`a della Calabria, Rende, Italy\\
    $^{4}$Southwest Research Institute, Boulder, CO, USA\\
    $^{5}$National Center for Atmospheric Research / High Altitude Observatory, Boulder, CO, USA\\
    $^{6}$NASA Goddard Space Flight Center, Greenbelt, MD, USA\\
    $^{7}$Naval Research Laboratory, Washington, DC, USA\\
    $^{8}$Southwest Research Institute, San Antonio, TX, USA\\
    $^{9}$United Kingdom Research \& Innovation / Science \& Technology Facilities Council / Rutherford Appleton Laboratory, Harwell Campus, Oxfordshire, U.K.
}

\date{Accepted XXX. Received YYY; in original form ZZZ}

\pubyear{2026}

\begin{document}
\label{firstpage}
\pagerange{\pageref{firstpage}--\pageref{lastpage}}
\maketitle

\begin{abstract}
We investigate the radial evolution of solar wind turbulence using remote-sensing brightness observations from the Solar Terrestrial Relations Observatory (STEREO) and the first science-ready images from the Polarimeter to Unify the Corona and Heliosphere (PUNCH) mission. Exploiting the combined wide fields of view of STEREO's COR2 coronagraph (4–9 \rsun) and PUNCH's Wide Field Imager (WFI, 21–50 \rsun), we track the radial evolution of the power spectral density and correlation scale of density fluctuations from 4 to 50 \rsun. Line-of-sight integration systematically steepens the ``usual'' turbulence scaling laws by a constant offset of about $-0.5$ across all resolved scales, so that the energy-containing range (nominally $k^{-1}$) and the inertial range (nominally $k^{-5/3}$) appear as $k^{-1.5}$ and $k^{-2.1}$, respectively -- consistent with previous synthetic-image simulations. The correlation scale increases with heliocentric distance across the full range of heliodistances probed, in accordance with previous theoretical models and observations at closer distances from the Sun.
\end{abstract}

\begin{keywords}
keyword1 -- keyword2 -- keyword3
\end{keywords}

\section{Introduction}
The continuous expansion of the solar wind into the heliosphere is mediated by large-scale flows and more local magnetohydrodynamic (MHD) turbulence. The latter drives the transport of energy from large, macroscopic scales down to kinetic dissipation scales, ultimately heating the coronal and heliospheric plasma. Understanding the dynamical development of the solar wind and its spatiotemporal evolution remains one of the most prominent problems in heliophysics \citep{brunocarbone2005review,matthaeus2011who}.

Studies of heliospheric dynamics have been conducted based on a persistent dichotomy between in-situ and remote-sensing paradigms. Both approaches have progressed in parallel, each offering a fundamentally different perspective on the expanding plasma. In-situ measurements -- from missions like Helios, Wind, Parker Solar Probe (PSP), and Solar Orbiter -- provide high-cadence local time series of magnetic field and plasma moments. However, despite the high level of detail, point-like measurements suffer from a localized spatial bias. In-situ sensors are restricted to one-dimensional, thread-like trajectories through a three-dimensional, highly structured, and dynamically expanding medium. Consequently, in-situ observations are subject to space-time aliasing, making it notoriously difficult to separate spatial structures from temporal variations \citep{taylor1938spectrum_frozenin,matthaeus2016ensemble,pecora2025spacetime}. Previous attempts to measure turbulence evolution in situ \citep{chen2020evolution} compared samples across vast heliocentric distances. These comparisons are often compromised by non-uniform solar wind conditions or must rely on rare, nearly radial alignments of multiple spacecraft \citep{telloni2021evolution}, making it exceptionally challenging to map a single plasma parcel's continuous evolution. Analogous analyses have been conducted using Cluster \citep{escoubet1997cluster} and Magnetospheric Multiscale \citep[MMS, ][]{burch2016MMS} mission data in the terrestrial magnetosheath, where spatial evolution of spectral properties was also observed \citep{yordanova2008magnetosheath,li2020evolution,pecora2026general_ubmsh}.

Wide-field remote-sensing imagers offer a complementary perspective by capturing two-dimensional spatial maps of the solar wind. However, heliospheric imagers and coronagraphs like COR2 from the Solar Terrestrial Relations Observatory \citep[STEREO;][]{kaiser2008STEREO}, and the Wide-Field Imager for Solar Probe Plus onboard PSP \citep[WISPR; ][]{vourlidas2016WISPR} are restricted by a limited and often fragmented field of view, which prevents continuous tracking from the lower corona out to 1~au. More fundamentally, remote-sensing instruments generally do not record local, point-like properties. Instead, they capture sunlight that has been Thomson-scattered by free electrons, resulting in a column-integrated measurement of brightness along a line of sight  \citep[LOS;][]{howard2012thomson,deforest2013thomson}. This projection acts as a non-trivial spatial filter that superposes plasma volumes of different densities, mixes different orientations of the local magnetic field, and attenuates small-scale structures.

This line-of-sight integration systematically modifies the properties of the solar wind's turbulence spectra. Early remote-sensing analyses of density fluctuations, such as the azimuthal power spectra extracted from STEREO data by \citet{deforest2018highly}, revealed deviations from typical in-situ scaling laws within the energy-containing range. Rather than indicating physically different nonlinear processes, these deviations represent the intrinsic consequence of the LOS projection. In this work, we present an empirical framework to interpret these integrated spectral observations.
Section~\ref{sec:math} mathematically details why LOS integration shifts the perceived spectral indices, building on previous numerical verification \citep{pecora2024mhdPUNCH}.
Section~\ref{sec:obs} presents the observational results from STEREO-A/COR2 and first data from the Wide Field Imager \citep[WFI,][]{laurent2026wfipunch} onboard the Polarimeter to Unify the Corona and Heliosphere \citep[PUNCH,][]{deforest2026PUNCH} mission that provide unprecedented images for deriving the radial evolution of angularly averaged spectra. Leveraging the integration of multiple wide-field polarimetric imagers into a continuous, wide-angle field of view, PUNCH provides the observational continuity required to track the radial evolution of these spectra.
Section~\ref{sec:disc} presents our summary and discussion of the results.

\section{In-situ vs. Remote-Sensing Spectra}
\label{sec:math}
The power spectral density of turbulent fields provides basic information on the transfer of energy from large to small scales. In homogeneous, isotropic hydrodynamic turbulence, the energy provided to the system at some rate produces eddies of about the system size. These, due to nonlinear interactions, fragment into progressively smaller eddies, producing a so-called ``cascade of energy'' from large to small scales. This cascade continues until molecular viscosity scales are reached and energy is dissipated. Magnetohydrodynamic systems, like the solar wind, have a similar behavior. This cascade of energy is observed in the power spectral density with a constant power-law scaling in the intermediate (inertial) range of scales between injection and dissipation \citep{MoninYaglomI,MoninYaglomII,frisch1995turbulence,pope2000turbulent}.
Here, we focus on how line-of-sight integration distorts the underlying spectra. Let us contrast the formal definitions of in-situ and remote-sensing power spectra. Consider a local fluctuating physical field $f(\mathbf{x})$. For the present context, it can be \comm{the electron density $n$} that is proportional to local contributions to image brightness. An in-situ spacecraft traversing the plasma measures the local instantaneous values of this field, recording them as a time series. Assuming Taylor's hypothesis \citep{taylor1938spectrum_frozenin}, time series can be converted into spatial information by multiplying the temporal lag by the solar wind speed.
Assuming homogeneity of fluctuations, the spatial two-point autocorrelation function is defined as 
$R(\mathbf{l}) = \langle f(\mathbf{x}) f(\mathbf{x} + \mathbf{l}) \rangle$
where $\mathbf{l} = l \hat{\mathbf{e}}_x$ represents the spatial separation (or lag) along the trajectory $\hat{\mathbf{e}}_x$, and $\langle \cdot \rangle$ denotes an ensemble or spatial average.
In the region over which statistical homogeneity remains valid, the autocorrelation function 
$R(\mathbf{l})$ is independent of absolute position $\bf x$ \citep{batchelor1953theory}.
\comm{The one-dimensional, longitudinal, in-situ power spectral density (PSD), $S(k)$ with wavenumber $k=2\pi/l$, is defined as the Fourier transform of the autocorrelation function }\citep{wiener1930generalized,blackman1958measurementI,blackman1958measurementII} $S(k) = \int_{-\infty}^{\infty} dl \, e^{-ikl} R(l,0,0)$,
where we (arbitrarily) choose the $\mathbf{x}$ direction to align with the trajectory.
For classic Kolmogorov-like turbulence in the inertial range, $S(k) \propto k^{-5/3}$, while for the low-frequency, energy-containing range, we often observe $S(k) \propto k^{-1}$ \citep{brunocarbone2005review,kiyani2015dissipation,wang2024noisePUNCH}.

Remote-sensing instruments do not sample the local field $f(\mathbf{x})$. Instead, they measure an integrated quantity $\bar{f}(r_{\mathrm{pos}})$.
For simplicity, in the rest of this treatment, we adopt the small-angle approximation and treat this operation as direct orthographic projection onto a sky plane perpendicular to the Sun--Earth line, via integration along LOS parallel to the Cartesian X axis; the results are understood to generalize to the wide-angle PUNCH field of view. For a coordinate system where $\ell$ represents the distance along the LOS and \comm{$r_\mathrm{pos} $ is the distance from the Sun center to a LOS in the plane of sky (YZ)} \citep[for a complete diagram of the geometry, see Fig.~18 in][]{deforest2026PUNCH}, the observed field is $\bar{f}(\mathrm{Y, Z}) = \int d\bm{\ell} \ w(\bm\ell) f(\bm\ell)$, where $w(\bm\ell)$ is a spatial weighting function including effects along the LOS integration path $\bm\ell$ such as the Thomson scattering geometry and the $\propto r^{-2}$ radial fall-off of the solar wind density.
Without loss of generality, we can write the projected two-point autocorrelation function calculated in the sky plane (Y, Z) with increments $l$ in the Y direction as
$R'(l) = \langle \bar{f}(\mathrm{Y, Z}) \bar{f}(\mathrm{Y} + l, \mathrm{Z}) \rangle$.
The resulting remote-sensing power spectrum $S'(k)$ is the Fourier transform of this projected autocorrelation $ S'(k) = \int_{-\infty}^{\infty} dl \, e^{-ikl} R'(l) \neq S(k).$
$S'(k)$ is an integral of the spatial correlation function of the weighted field obtained from LOS integration, mathematically distinct from the 1D in-situ spectrum $S(k)$. The LOS integration generally acts as a low-pass spatial filter along the $\ell$ direction, reducing high-frequency fluctuations and further altering the apparent spectral indices of the turbulence regimes.

Computing the spatial power spectrum $S'(k)$ directly via Fourier transforms requires considering several caveats. Due to the finite field of view of remote sensors, non-periodic spatial boundaries, data gaps, or coordinate transformations in the image plane, direct Fourier transforms often suffer from windowing artifacts, spectral leakage, and numerical instability. To circumvent these issues, we compute an equivalent power spectrum from the second-order structure function \citep{chasapis2017high,pecora2024mhdPUNCH,bishop2026direct}.
This approach partially alleviates some of these problems due to the more rapid convergence of the structure function and its reduced sensitivity to weak inhomogeneity \citep{pugachev1965theory}.
\comm{Without loss of generality, we can assume that the previously mentioned field $f(\mathbf{x})$ is 1-dimensional -- thus we can write the generic position $\mathbf{x}$ as $x$. Consequently, the second-order structure function is defined as:}
\begin{equation}
    S^2(l) = \langle \left| f(x+l) - f(x) \right|^2 \rangle.
    \label{eq:s2}
\end{equation}
For a homogeneous turbulence system where the power spectrum scales as $S(k) \propto k^{-\alpha}$ (with $1 < \alpha < 3$), the structure function behaves as $S^2(l) \propto l^{\alpha - 1}$ \citep{frisch1995turbulence}. Thus, by mapping the spatial dimension as $k \mapsto 1/l$, we can construct the equivalent power spectrum as
\comm{
\begin{equation}
    S_\mathrm{eq}(1/l) \equiv l\,S^2(l).
    \label{eq:eq_spec}
\end{equation}
Thus, the equivalent spectrum scales as $l^{\alpha}$ or, equivalently, $k^{-\alpha}$, preserving the identical spectral index $\alpha$ of the underlying turbulent cascade while bypassing the sensitivity of standard Fourier transforms to window edge effects, yielding a robust and highly stable spectral estimator for both synthetic and observational wide-field data.}

\textbf{Energy-containing range --}
As mentioned earlier, the range of scales beyond a typical correlation length shows a power spectral density that scales as $k^{-1}$. This can be due to several mechanisms, such as coalescence of flux tubes \citep{matthaeus1986low}, or superposition of different solar wind streams with log-normally distributed correlation lengths \citep{machlup1981earthquakes,montroll1982onef,pradata2025observations_apjl_mercury1f,wang2026fnoise_superposition}.
By using remote observations, this range of scales was previously analyzed by \cite{deforest2018highly} using a deep-exposure campaign from STEREO/COR2. The resulting spectra were found to have a slope proportional to $\sim k^{-1.5}$, expected to correspond to in-situ $k^{-1}$ after consideration of the LOS integration effects.

\textbf{Inertial range --} At higher wavenumber, the energy-containing range transitions toward the inertial range where nonlinear interactions dominate the dynamics. Here, for homogeneous turbulence systems, the PSD scales as $k^{-5/3}$ \citep{frisch1995turbulence,pope2000turbulent}. Some studies argue that solar wind observations show a transition from $k^{-3/2}$ closer to the Sun to $k^{-5/3}$ near 1~au \citep{chen2020evolution}. 
We do not address this distinction here, as current remote sensing resolutions may be insufficient to provide a conclusive result; we defer this refinement to future work.
Our approach is to compare with previous numerical results that used the FORWARD-modeling synthesis of white-light data \citep{gibson2016forward} to replicate the observations of PUNCH-like instruments within an MHD simulation \citep{pecora2024mhdPUNCH}. The results showed a net deviation from a Kolmogorov-like ($-5/3$) spectrum in the inertial range of the simulation to an apparent slope that was slightly steeper than $-2$ in the synthesized white-light image.

\section{Remote-Sensing Observations of Solar Wind Spectral Properties}
\label{sec:obs}
In this section, we present some results obtained from two generations of remote-sensing missions: STEREO-A/COR2 \citep{kaiser2008STEREO} and PUNCH/WFI \citep{laurent2026wfipunch}.

\subsection{STEREO Observations}

We used image data from the STEREO-A/COR2 coronagraph \citep{howard2008secchi}. Level-0.5 data files containing total brightness were downloaded from \url{https://stereo-ssc.nascom.nasa.gov/data/ins_data/secchi/L0/a/img/}. We chose observations of a quiet period of minimum solar activity (2007 August 6, 04:07:30 UT) to ensure that transient phenomena, such as coronal mass ejections, do not affect our analysis. IDL SolarSoft \citep{freeland1998data} routines made available by the STEREO instrument team (\url{https://soho.nascom.nasa.gov/solarsoft/stereo/secchi/doc/sccuserguide.html}) were used to process and calibrate the data. The F-coronal background was removed by dividing by a monthly-minimum image rather than by subtraction; as discussed below, this choice modifies the absolute radial amplitude normalization of the brightness fluctuations but is necessary to adequately suppress residual artifacts for an accurate analysis of spectral slopes.

\begin{figure}[ht]
    \centering
    \includegraphics[width=0.9\linewidth]{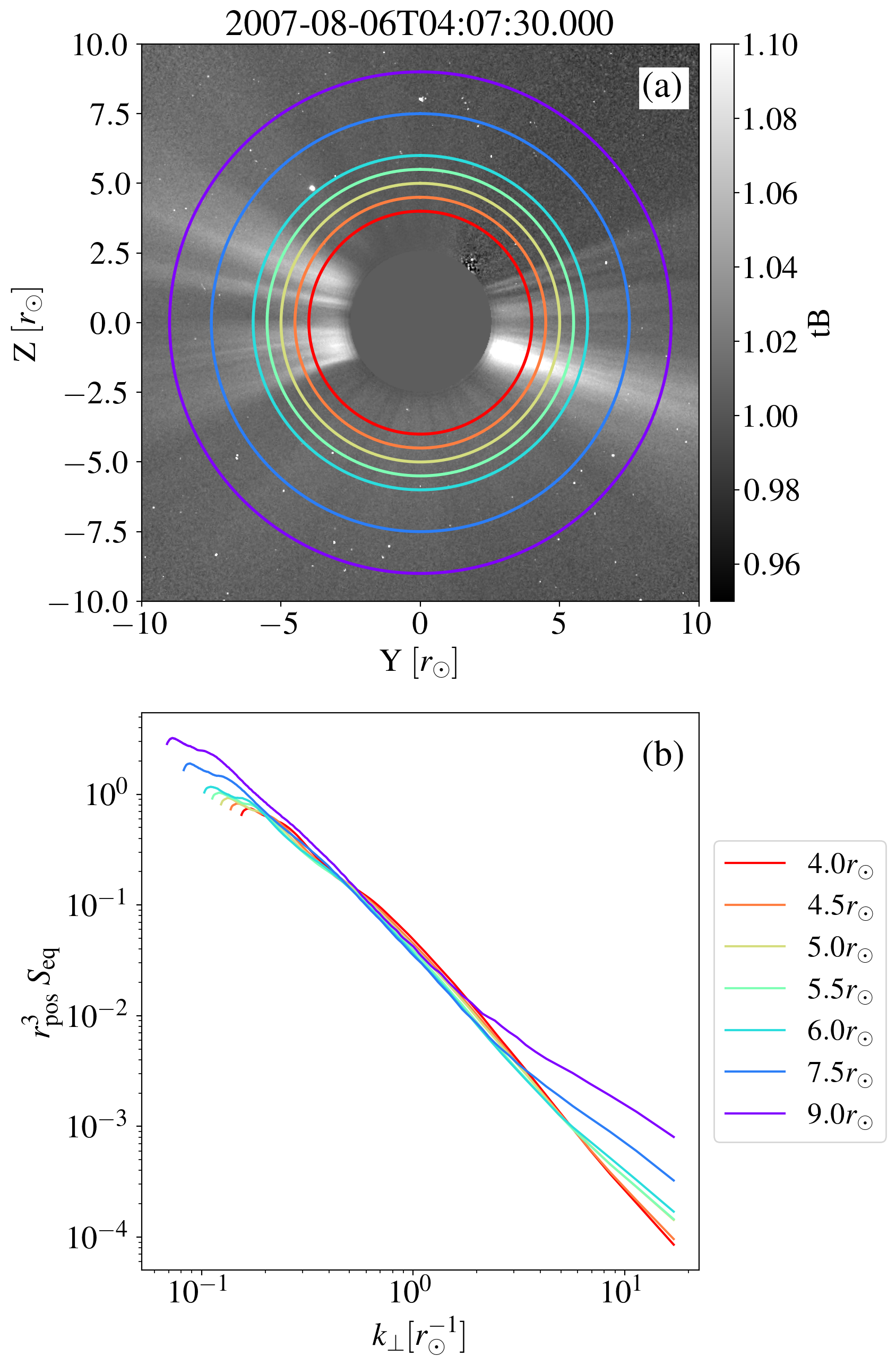}
    \caption{(a) STEREO-A/COR2 total brightness on 2007 Aug 6, 04:07:30 UT.
    F-corona removed by dividing by a monthly-minimum background image (see Section 3.1); this affects the absolute amplitude scaling shown here but not the derived spectral slopes. (b) Equivalent spectra calculated using information from different concentric circles in the azimuthal direction, same color as in (a). Spectra are rescaled by heliodistance in the plane of the sky to the power of 3.}
    \label{fig:stereo1}
\end{figure}

\comm{Structure functions are evaluated along concentric circles of radius $r_{\mathrm{pos}} = \sqrt{Y^2 + Z^2}$ in the sky plane. For any pair of points $(Y_1, Z_1)$ and $(Y_2, Z_2)$ located on the same circumference, the spatial separation $l$ is calculated as the Cartesian Euclidean distance (chord length) $l = \sqrt{(Y_2 - Y_1)^2 + (Z_2 - Z_1)^2}$. Equivalent spectra are then obtained via Eq.~\ref{eq:eq_spec}, with the spatial wavenumber (perpendicular to the radial expansion direction) mapped as $k_{\perp} \sim 1/l$.}
Specifically, we focus on increasing heliocentric distances from 4~\rsun to 9~\rsun, as illustrated in Fig.~\ref{fig:stereo1}(a). We avoid distances where the presence of stars and other artifacts in the background can impact the analysis. The resulting equivalent spectra are shown in Fig.~\ref{fig:stereo1}(b).
Overall, as expected, the amplitude of the brightness fluctuations decreases with heliocentric distance. Brightness is proportional to the column-integrated electron density, whose radial trend in the heliosphere is well established from in-situ observations \citep{gazis1983radial,hellinger2011heating,perrone2019radial}.
Since the relative fluctuation amplitude $\delta n/n$ is known to remain approximately constant with heliocentric distance \citep{roberts1987nature}, the density fluctuations $\delta n$ are expected to follow the same radial scaling as the background density itself.
We report scaled spectra where the scaling factor has been determined empirically to obtain a good overlap. In this range of heliodistances, the brightness fluctuations squared are found to overlap when scaled as $\sim r_{\rm pos}^{-3}$.
We note that this empirical exponent is shallower than the physically expected value (closer to $\sim r_{\rm pos}^{-6}$) because the division-based F-corona removal used in the STEREO pipeline systematically distorts the absolute amplitude normalization with heliocentric distance. Because this distortion affects only the overall amplitude and not the shape of the spectra, it does not compromise the spectral-slope analysis or the estimates of the correlation scale that follow.

\begin{figure}[ht]
    \centering
    \includegraphics[width=0.9\linewidth]{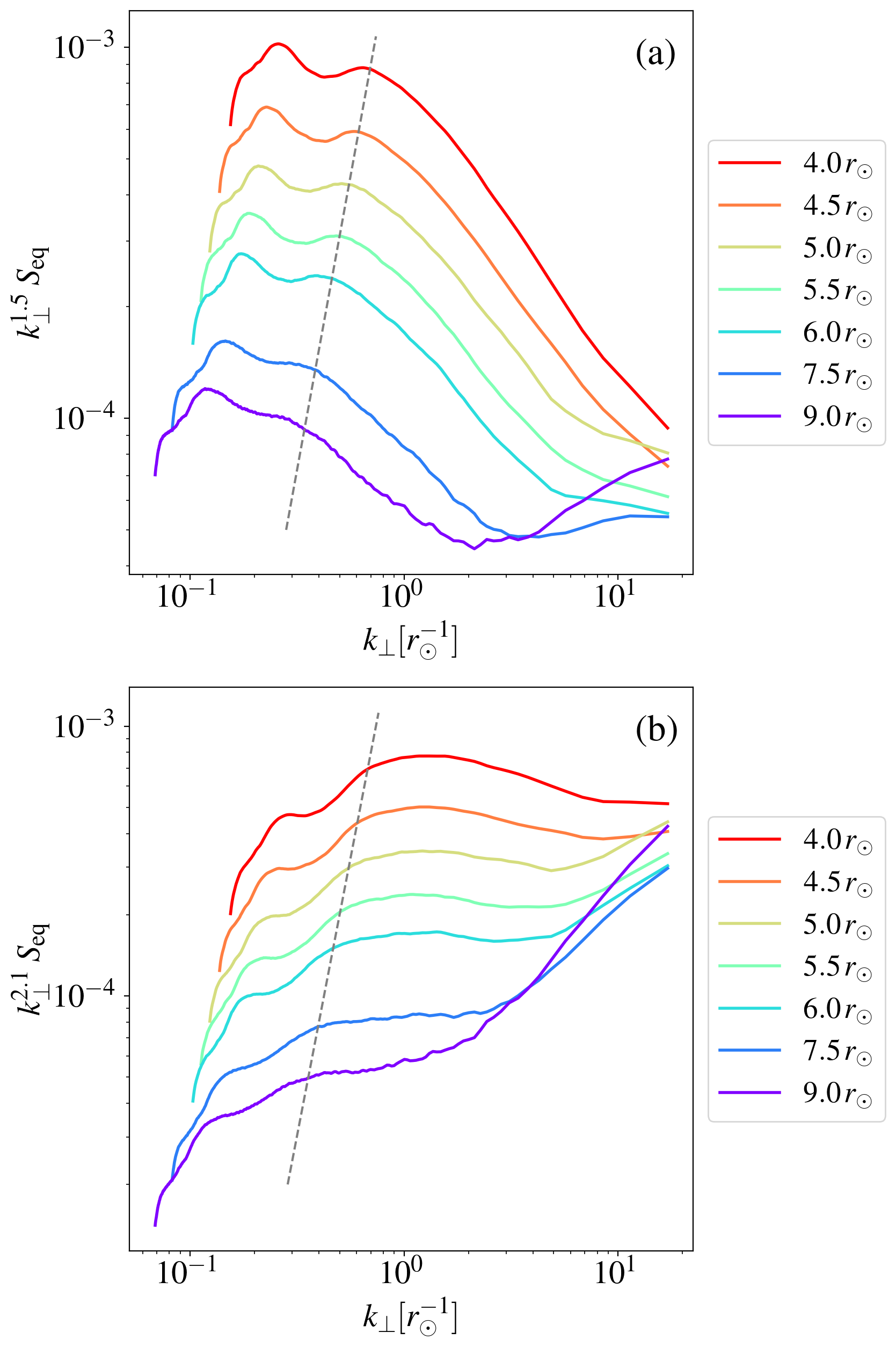}
    \caption{Equivalent spectra compensated by (a) $k_\perp^{1.5}$ for the energy-containing and by (b) $k_\perp^{2.1}$ for the inertial ranges, computed from STEREO/COR2 observations (Fig. \ref{fig:stereo1}). The oblique dashed lines are indicative of the spectral break (interruption of the flat-ish regions).}
    \label{fig:stereo2}
\end{figure}

The range of scales in which a specific spectral slope is attained can be visualized more clearly via the use of the so-called compensated spectra. Specifically, the spectra are multiplied by the (inverse) power of the wavenumber corresponding to the expected slope, so that a relatively flat region directly highlights the range of scales where such a slope is established. In Fig.~\ref{fig:stereo2}(a), we present the equivalent spectra compensated by \comm{$k_\perp^{1.5}$} to emphasize the energy-containing range to the left of the oblique dashed line, whereas in Fig.~\ref{fig:stereo2}(b), the spectra are compensated by $k_\perp^{2.1}$, which reveals the emergence of an inertial range to the right of the oblique line.
As mentioned above, we refer to these scale ranges as energy-containing and inertial in deference to the fact that their slopes can be mapped to the ``usual'' $1/f$ and Kolmogorov's scalings, respectively, as already evidenced by previous observational \citep{deforest2018highly} and numerical \citep{pecora2024mhdPUNCH} works. However, since the LOS integration operates as a low-pass filter, the direct mapping of such scale ranges to the actual in-situ numerical values is not necessarily trivial.

Several interesting features become visible by inspecting the compensated spectra. In Fig.~\ref{fig:stereo2}(a), it can be observed that, as the heliocentric distance increases, the range of scales pertaining to the energy-containing range systematically shifts towards smaller $k$ values, a behavior that can be attributed to the solar wind expansion.
A similar evolution of the inertial range is neatly discernible in Fig.~\ref{fig:stereo2}(b). Here, we notice that the onset of the inertial range appears at progressively smaller $k$ values, possibly a direct consequence of solar wind expansion and the dynamics of the turbulence cascade. These, and possibly other effects, cause a shift of the energy-containing range toward larger scales \citep{brunocarbone2005review}, as has been observed in remote-sensing data during total solar eclipse ranging from the chromosphere to the coronal base \citep{bailey2025measurement} as well as in in-situ measurements in the outer corona \citep{cuesta2022isotropization}.

\subsection{PUNCH Observations}

\begin{figure}
    \centering
    \includegraphics[width=0.9\linewidth]{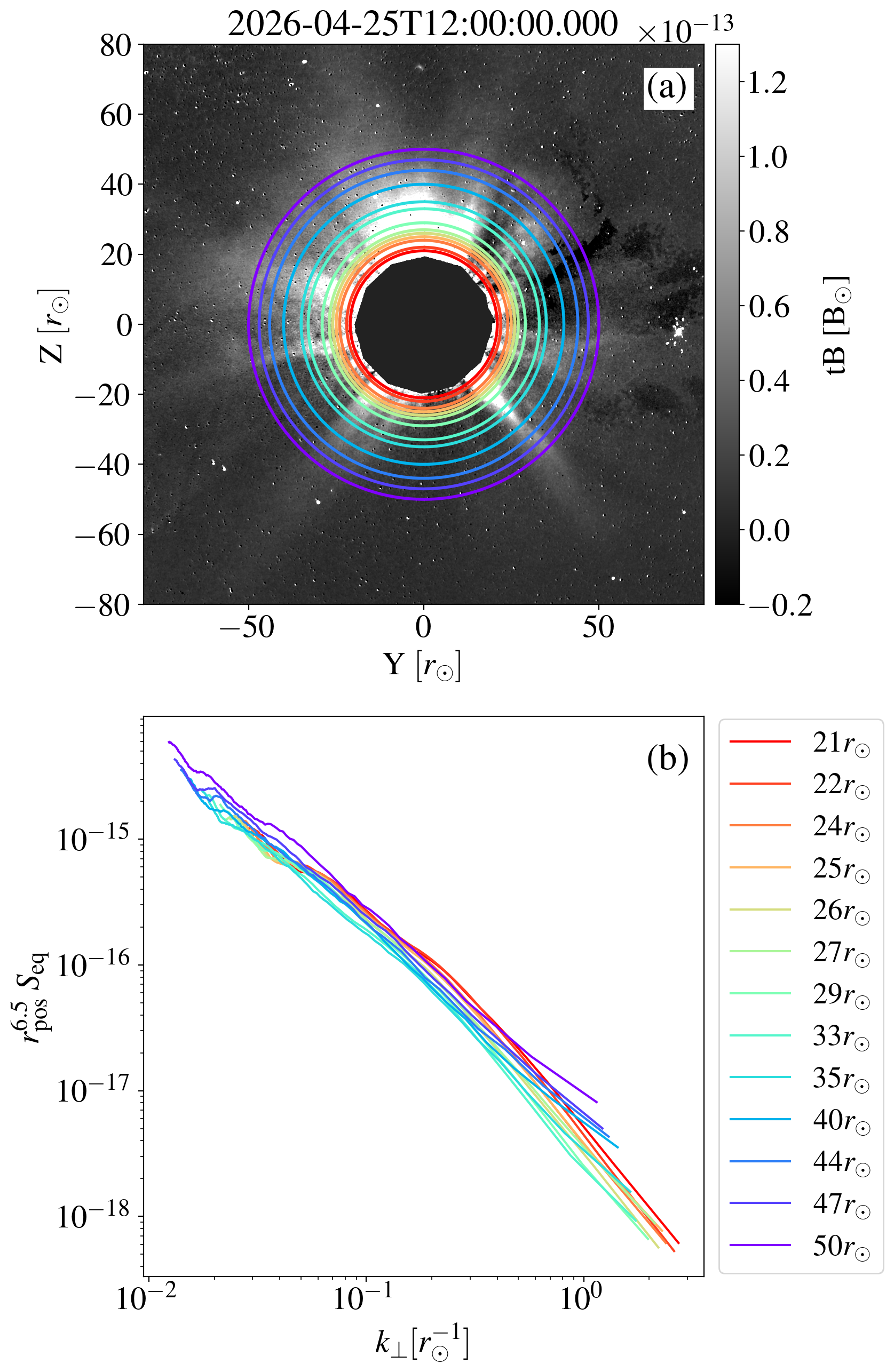}
    \caption{(a) PUNCH WFI total brightness (in units of mean solar brightness, B$_\odot$) image on 2026 Apr 25, 12:00 UT. (b) Equivalent spectra whose intensities overlap when scaled by $r_\mathrm{pos}^{6.5}$. Each line corresponds to the same-colored circle in (a).}
    \label{fig:punch1}
\end{figure}

The PUNCH mission comprises four synchronized satellites operating as a single virtual instrument. Together, they provide continuous, wide-field, polarized imaging from $1.5^\circ$ to $45^\circ$ (roughly $6\,r_\odot$ to $180\,r_\odot$), delivering unprecedented full angular coverage across this immense field of view \citep{deforest2026PUNCH}.
Specifically, the Narrow Field Imager (NFI) \citep{colaninno2025punchNFI} is designed to observe the inner region from $1.5^\circ$ to $9^\circ$ (6 \rsun to 36 \rsun) at a spatial resolution of $0.5$ arcmin per pixel, 
while the three Wide Field Imagers (WFIs) \citep{laurent2026wfipunch} cover the outer region from $5^\circ$ to $45^\circ$ (20 \rsun to 180 \rsun) at a resolution of $1.5$ arcmin per pixel.
However, because NFI data is not yet available, we used STEREO/COR2 observations to provide coverage for the regions closer to the Sun. PUNCH observations complement those from STEREO, extending the coverage outward into the inner heliosphere, as can be appreciated by noting the spatial extent in the $Y$ and $Z$ directions in Figs.~\ref{fig:stereo1} and \ref{fig:punch1}.
\comm{The calibration pipelines for STEREO and PUNCH handle F-corona removal differently: STEREO normalizes the background, whereas PUNCH subtracts it. Consequently, raw amplitude levels cannot be directly compared between the two data sets. However, this methodological difference does not affect the spectral properties analyzed below.}
\comm{Although the COR2 field of view extends out to $15\,r_{\odot}$, we limited our COR2 analysis to $9\,r_{\odot}$. Beyond this distance, a reduction in the signal-to-noise ratio prevented the clear identification of spectral plateaus.}

The analyses for the PUNCH mission are conducted during a quiet period centered on 2026 April 25 at 12:00. Our investigation utilizes Clear Averaged Mosaic data (CAM, Level 3) version v0k, which provide the entire circular field of view by averaging over 32 minutes. These are the first available science-ready data products from the mission, and have the background F-corona removed and the starfield suppressed. By virtue of the strong complementarity between STEREO and PUNCH, and thanks to the significantly extended field of view provided by the latter, we are able to expand the analysis of STEREO observations in the previous section, limited to the range between 4 and 9 \rsun, out to a distance of 50 \rsun.
Using the same methodology, we define a series of azimuthal concentric rings along which the total brightness is sampled. From these observational data, we computed the second-order structure function and the corresponding equivalent spectra, which are illustrated in Fig.~\ref{fig:punch1}.
In contrast to the $\sim r_{\rm pos}^{-3}$ amplitude scaling recovered from STEREO Fig.~\ref{fig:stereo1}(b), which -- as discussed in Section 3.1 -- is suppressed by the division-based F-corona removal, the squared radiance from PUNCH (which does not suffer from this artifact) scales as $\sim r_{\rm pos}^{-6.5}$, in agreement with previous results \citep{deforest2016fading,deforest2018highly}.

\begin{figure}
    \centering
    \includegraphics[width=0.9\linewidth]{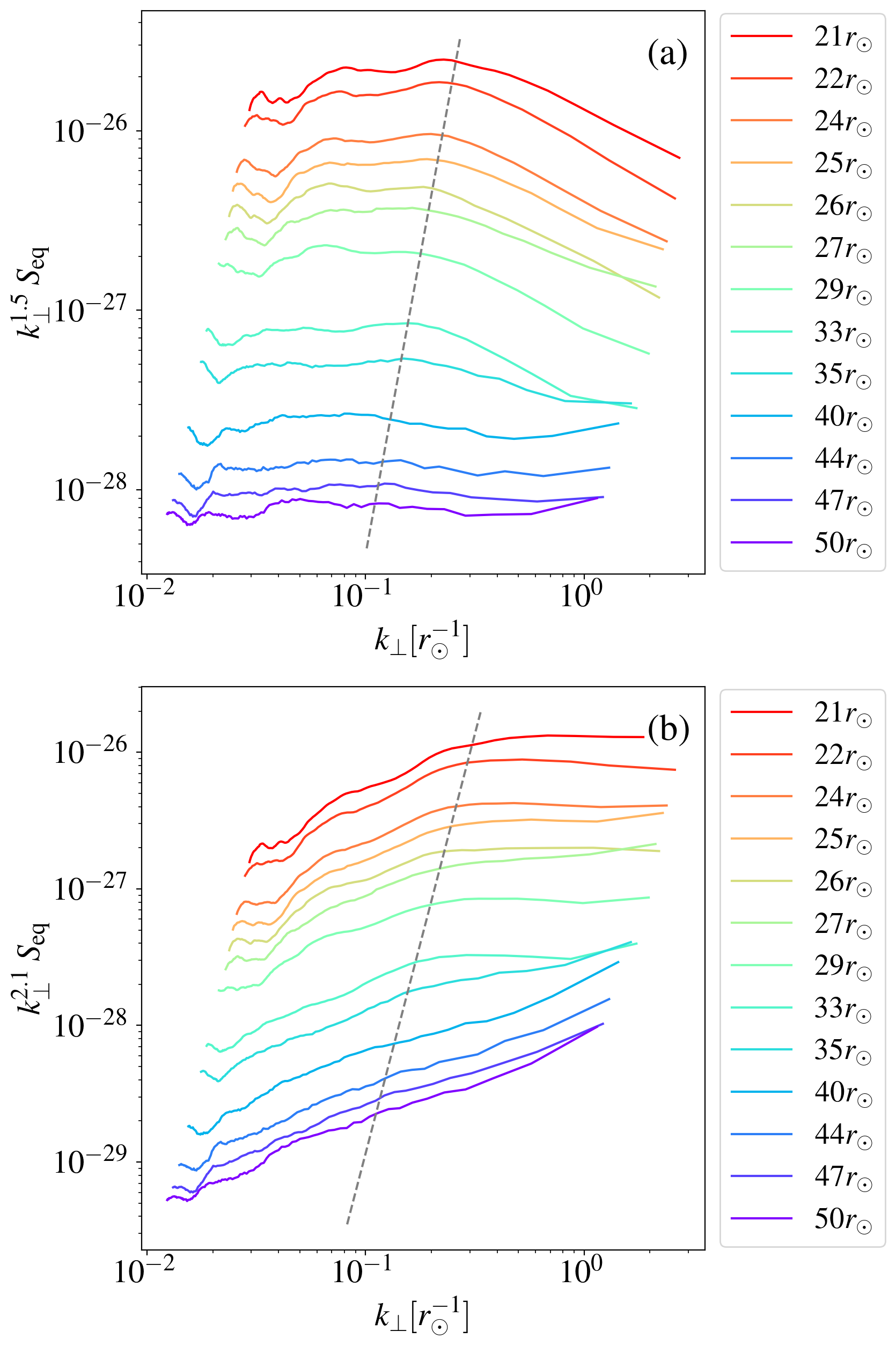}
    \caption{Equivalent spectra compensated by (a) $k_\perp^{1.5}$ for the energy-containing and by (b) $k_\perp^{2.1}$ for the inertial ranges, computed from PUNCH/WFI observations (Fig. \ref{fig:punch1}). The oblique dashed lines are indicative of the spectral break (interruption of the flat-ish regions).}
    \label{fig:punch2}
\end{figure}

To visualize the possible presence of energy-containing and inertial ranges, we plot the compensated spectra in Fig.~\ref{fig:punch2}. The relatively horizontal regions in panel (a) to the left of the oblique dashed line illustrate the spatial scales pertinent to the energy-containing range, whereas the inertial range emerges as the flat region in panel (b) to the right of the oblique line.

Empirically, we observe that across all investigated scales, from the energy-containing to the inertial range, the power spectral density scaling undergoes a uniform steepening by a constant value of approximately $-0.5$ with respect to expected in-situ observations. Consequently, the characteristic spectral slopes corresponding to the energy-containing range ($k^{-1}$) and the inertial range ($k^{-5/3}$) are both shifted by the same amount, yielding modified exponents of $k^{-1.5}$ and $k^{-2.1}$, respectively. Although a comprehensive analytical explanation remains challenging at present, these observational findings are entirely consistent with the results obtained from numerical simulations. Specifically, this steepening effect emerges clearly when comparing the intrinsic MHD density field -- which displays a standard Kolmogorov-like spectrum -- with the synthetic PUNCH-like images produced via the FORWARD modeling tool \citep{gibson2016forward}, which instead exhibit a noticeably steeper spectrum with a slope slightly below $-2$ \citep{pecora2024mhdPUNCH}.

\begin{figure}[ht]
    \centering
    \includegraphics[width=0.9\linewidth]{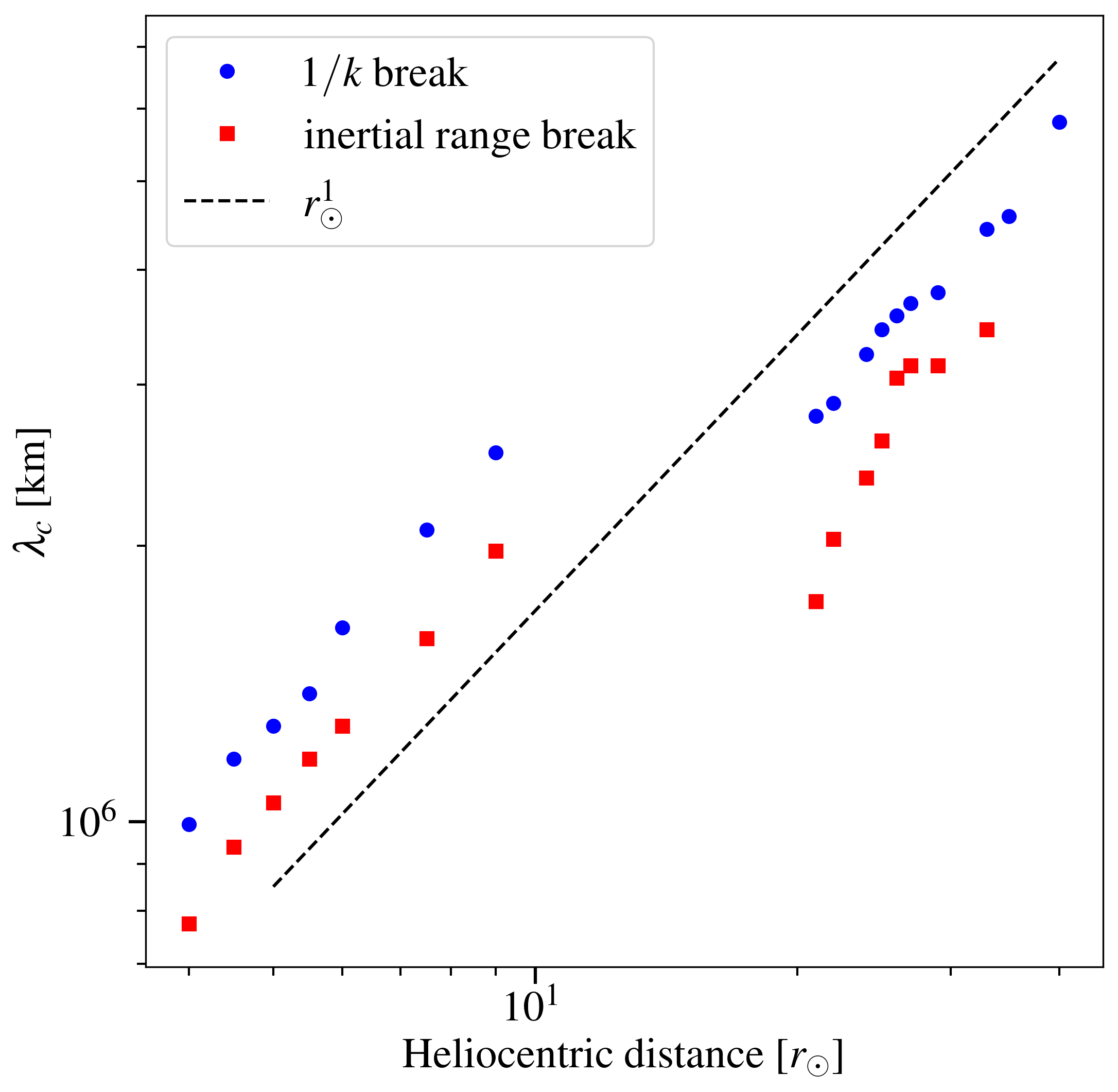}
    \caption{Evolution of the correlation length $\lambda_c$ as a function of heliodistance measured as inverse of the spectral breaks (oblique dashed lines in Figs.~\ref{fig:stereo2} and \ref{fig:punch2}) of the $1/k$ range (blue circles) and of the inertial range (red squares), for STEREO (4 to 9 \rsun) and PUNCH (21 to 50 \rsun). The dashed line indicates a linear increase in the correlation scale with distance from the Sun.}
    \label{fig:corrlen}
\end{figure}

Finally, in order to investigate the radial evolution of the correlation scale $\lambda_c$, we evaluate the variation of the spectral break scale separating the energy-containing range from the inertial range as a function of the heliocentric distance.
The correlation scale can be roughly estimated as the spatial scale corresponding to the beginning of the inertial range, where it connects to the energy-containing range. We measure the specific $k$ values corresponding to the spectral break (identified visually) for both the energy-containing and the inertial ranges (indicated by oblique dashed lines in Figs.~\ref{fig:stereo2} and \ref{fig:punch2}). Namely, we identify the $k$ values corresponding to the bendover of the energy-containing range proceeding towards larger $k$ values, as well as the $k$ values corresponding to the bendover of the inertial range moving from larger to smaller $k$ values. The inverse of these spectral break $k$ values as a function of the heliocentric distance is reported in Fig.~\ref{fig:corrlen}.
We estimated the breaks separately, going from small to large $k$'s in the energy-containing range, and from large to small $k$'s in the inertial range to avoid the in-between scales where the PSD bends continuously to transition from one slope to the other. However, there is little difference between these two scales, as seen in Fig.~\ref{fig:corrlen}.
\comm{While Figs.~\ref{fig:punch1} and \ref{fig:punch2} display spectra across 13 radial distances to show the continuous radial trend, Fig.~\ref{fig:corrlen} reports the spectral break scale for a subset of 8 distances. Specifically, we included only those heliocentric distances where both the energy-containing and inertial ranges exhibited clear plateaus in the compensated spectra. Distances where the spectral transition was broader or degraded by noise were omitted from the break tracking to ensure accuracy.}

The correlation (break-point) scale exhibits a roughly linear increase with heliocentric distance across the entire range spanned by our observations, from $4\,r_\odot{}$ to $50\,r_\odot$. A discontinuity of approximately a factor of two is present between the STEREO and PUNCH estimates: extrapolating the linear trend derived from STEREO to a distance of $20\,r_\odot$ predicts a correlation scale of $\lambda_c \sim 5\times10^{6}$~km, whereas PUNCH measures $\lambda_c \sim 2\times10^{6}$~km at the same heliocentric distance. We discuss the possible physical and instrumental origin of this offset, together with the broader interpretation of the correlation-scale evolution, in the discussion section.

\begin{figure}[ht]
    \centering
    \includegraphics[width=0.9\linewidth]{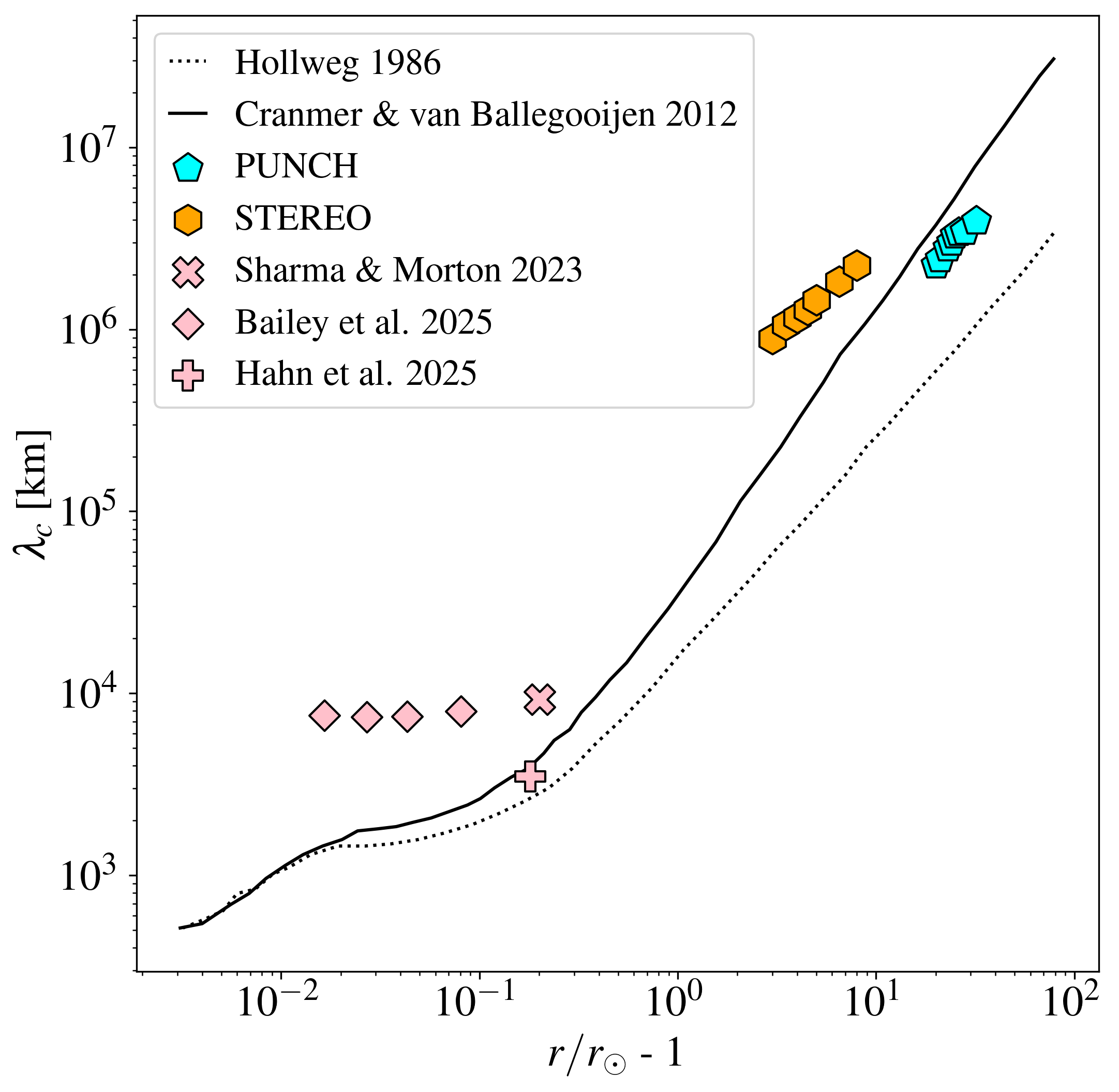}
    \caption{Comparison of correlation lengths with previous observational results and theoretical models. PUNCH (cyan pentagons) and STEREO (orange hexagons) data points are the average of the $1/k$ and inertial-range break scales from Fig.~\ref{fig:corrlen}. Other data points are taken from \citet{sharma2023transverse} (pink X, visible and near-infrared observations from CoMP and UCoMP), \citet{bailey2025measurement} (pink diamonds, white-light observations during a total solar eclipse), and \citet{hahn2025measurement} (pink plus, Fe~XIII 1074~nm line from DKIST). Theoretical models are shown for \citet{hollweg1986transition} (dotted line) and \citet{cranmer2012proton} (solid line).}
    \label{fig:corrlen_model}
\end{figure}

To place these correlation-length estimates in a broader context, Fig.~\ref{fig:corrlen_model} compares our STEREO and PUNCH values with a compilation of observational results and theoretical models spanning from the low corona out to several tens of solar radii. The STEREO and PUNCH data points shown here are computed as the average of the corresponding $1/k$ and inertial-range break scales reported in Fig.~\ref{fig:corrlen}. Alongside our measurements, we include the correlation lengths reported by \citet{sharma2023transverse}, \citet{bailey2025measurement}, and \citet{hahn2025measurement}, together with the theoretical models of \citet{hollweg1986transition} and \citet{cranmer2012proton}.
Our STEREO and PUNCH estimates extend to much greater heliocentric distances, offering a more extensive test of both theoretical models over roughly 1 to 50~\rsun. The \citet{cranmer2012proton} model provides the best match to our data, and it also reproduces the observations of \citet{sharma2023transverse} and \citet{hahn2025measurement} near 1~\rsun. The lower correlation length reported by \citet{hahn2025measurement}, relative to other observations at comparable distances, is consistent with the much higher spatial resolution of DKIST. In fact, finer resolution captures steeper gradients, which in turn causes the autocorrelation function to decay faster.
The measurements of \citet{bailey2025measurement}, on the other hand, sit systematically above both models. This offset can be accounted for by rescaling the models' correlation length at the coronal base, suggesting it may simply reflect a difference in baseline normalization. Alternatively, it could point to comparatively lower spatial resolution in that dataset, which -- following the same reasoning applied above to \citet{hahn2025measurement} -- would tend to inflate the inferred correlation length.

\section{Discussion and Conclusions}
\label{sec:disc}
In this work, we presented an investigation of the spectral evolution of solar wind density fluctuations utilizing remote-sensing observations from the STEREO/COR2 coronagraph and the first science-ready data from the PUNCH mission. A central theme of this analysis is the impact of LOS column-integration, which inherently acts as a non-trivial low-pass spatial filter along the observing direction: by superposing plasma volumes of varying densities and attenuating small-scale structures, this geometric projection systematically modifies the ``usual'' turbulence scaling laws observed via in-situ measurements or numerical simulations. 

Empirically, this integration produces a consistent spectral steepening of approximately $-0.5$ across all investigated spatial scales, in agreement with previous numerical predictions \citep{pecora2024mhdPUNCH}. As a result, the low-frequency energy-containing range, nominally $k^{-1}$ in situ, is perceived as $\sim k^{-1.5}$, while the inertial range steepens past the classic Kolmogorov ($k^{-5/3}$) scaling to a slope slightly steeper than $-2$. This is consistent with the steepening of the $1/k$ range reported by \citet{deforest2018highly}, where the energy-containing spectrum was found to steepen toward $-1.5$, and confirms that turbulence information extracted from remote-sensing observations follows a consistent, reproducible modification that favors synergistic cross-comparison of data from different instruments and astrophysical settings.
    
This reading contrasts with the interpretation of Solar Orbiter/METIS observations in \citet{telloni2024metis}, where slopes around $-1.5$ are attributed to the inertial range. Instead, we suggest that a slope of $-1.5$ from remote-sensing observations is consistent with the energy-containing range, and that the scales probed by \citet{telloni2024metis} are more likely to belong to that regime rather than the inertial one. We propose the same reinterpretation for the PSD reported by \citet{prete2025interaction} for X-ray observations of the supernova remnant SN 1006.

By exploiting PUNCH's wide field of view and its strong complementarity with STEREO, we mapped the evolution of power spectra from $4\,r_\odot{}$ to $50\,r_\odot$. As heliocentric distance increases, the characteristic scales delimiting the energy-containing range from the onset of the inertial cascade shift systematically toward smaller wavenumbers, a direct empirical signature of solar wind expansion.

\comm{The observed spectral features in Figs.~\ref{fig:stereo2} and \ref{fig:punch2} can be used to directly trace the dynamical evolution of solar structures in the expanding heliosphere. The energy-containing range ($k^{-1.5}$, corresponding to $k^{-1}$ in situ) reflects large-scale density and magnetic field inhomogeneities injected at the coronal base. Because non-linear interactions at these large scales are slow compared to the advection time, this range passively expands with heliocentric distance, preserving the spectral signature imprinted by the Sun. However, other relevant mechanisms such as coalescence of flux tubes and dynamical development are still debated \citep{matthaeus1986low,huang2023oneoverfPSP,wang2024noisePUNCH,wang2026fnoise_superposition}. 
Conversely, the emergence and broadening of the inertial range ($k^{-2.1}$, corresponding to in-situ Kolmogorov $k^{-5/3}$) with distance can be related to active MHD turbulent cascade driven by non-linear interactions. As the solar wind flows outward from 4 to 50 $r_{\odot}$, non-linear processes have sufficient time to erode progressively larger energy-containing structures, shifting the spectral break to lower wavenumbers. This continuous transition provides direct empirical evidence of the ``turbulent aging'' of solar wind plasma during its transit through the inner heliosphere \citep{matthaeus1998dynamical,chhiber2026dynamicalage}.}

Tracking the spectral break separating the two regimes, we find that the correlation scale $\lambda_c$ grows approximately linearly with heliocentric distance ($\lambda_c \propto r^{1}$) across the full range probed, from $4\,r_\odot{}$ to $50\,r_\odot$ (Fig.~\ref{fig:corrlen}). This is faster than the scalings typically reported by in-situ statistical surveys of the inner heliosphere: \citet{matthaeus1999correlation} found $\lambda_c \propto r^{0.67}$ between 1 and 10~au, and \citet{ruiz2014characterization} reported $\lambda_c \propto r^{0.43}$ between 0.3 and 5~au. It is, however, comparable to the result of \citet{cuesta2022isotropization}, who found a steeper $r^{1}$ scaling below 0.3~au ($\sim 65$ \rsun) that flattens to $r^{0.3}$ beyond it.

The absolute correlation-scale values recovered here are also several times larger than typical 1~au estimates, and roughly an order of magnitude larger than Parker Solar Probe measurements near $10\,r_\odot$ \citep{ruiz2014characterization,isaacs2015systematic,chhiber2021large,zhao2022turbulent,goodwill2026averaging}. We attribute this inflation to two complementary effects. First, LOS integration itself acts as a low-pass filter, suppressing the high-frequency fluctuations that would otherwise decorrelate the field more rapidly, which fictitiously broadens the estimated correlation scale. Second, our azimuthal sampling spans all latitudes and therefore includes the polar regions, where large, uniform coronal holes dominate during quiet periods. This is in contrast to in-situ observations, which are collected predominantly near the ecliptic plane and are continuously perturbed by the heliospheric current sheet, stream interaction regions, and (in the slow wind that dominates there) stronger velocity shears \citep{mccomas2003three_Ulysses}, all of which decorrelate fluctuations more quickly. Since the images analyzed here were chosen to avoid coronal mass ejections (CMEs) and other transients, and since the azimuthal sampling integrates over fast, uniform, largely undisturbed polar wind together with the ecliptic, our remote-sensing estimates are comparatively free of both effects, which plausibly explains the faster growth of $\lambda_c$ we recover relative to ecliptic in-situ surveys.

This contrast is compounded by a more general tendency of correlation-length estimates to grow with the size of the data sample: studies of longer interplanetary time series \citep{isaacs2015systematic} show correlation-scale estimates increasing with dataset length, an effect expected for stationary data whenever low-frequency, $1/f$-type power is present \citep{wang2024noisePUNCH}. Imaging analyses such as ours necessarily include these long-wavelength fluctuations, so correlation-scale estimates from imaging should generically exceed those from shorter, local in-situ records. An alternative correlation-scale definition that converges to a fixed value independent of record length has recently been proposed \citep{perez2026integral}; the integral scale obtained that way is about an order of magnitude larger than typical estimates, and potentially more compatible with the values reported here.

A further feature of our results is a systematic factor-of-two discrepancy between the two missions, with PUNCH (observations near solar maximum) yielding smaller correlation scales than STEREO (near solar minimum). This runs counter to the usual in-situ finding of larger magnetic-field correlation lengths near solar maximum \citep{wicks2008spatial,wicks2010variation,wang2025interplanetary_PNAS}, but those in-situ increases are largely driven by the higher frequency of CMEs and other large-scale transients near maximum. Because our images were specifically chosen to exclude CMEs, in order to isolate the pristine wind, the underlying physical differences in the ambient wind across the solar cycle can instead become the dominant effect: near solar minimum, coronal holes persist longer and cover a larger fraction of the solar surface in a relatively undisturbed state, providing coherent, large-scale structures that enhance the measured perpendicular correlation scale, consistent with the smaller PUNCH (solar-maximum) values.

Placing these correlation-length estimates in the broader observational and theoretical context reinforces this picture. Our STEREO and PUNCH values are broadly consistent with other observational determinations \citep{sharma2023transverse,bailey2025measurement,hahn2025measurement} and with the theoretical scalings of \citet{hollweg1986transition} and \citet{cranmer2012proton}, extending the empirical verification of both models to heliodistances of about 1--50 \rsun, with the \citet{cranmer2012proton} model tracking the combined dataset somewhat more closely across the whole range. The higher values reported by \citet{bailey2025measurement} can be reconciled by rescaling the models' correlation length at the coronal base, or may instead reflect the lower spatial resolution of those observations, which would tend to inflate the inferred correlation length.

Taken together, we regard the discrepancies in absolute correlation-length values -- between STEREO and PUNCH, and between remote-sensing and in-situ estimates more broadly -- as due to several geometric and dynamical effects: LOS integration, instrumental resolution, and the differential sampling of fast and slow wind populations across the solar cycle. The growth of the correlation scale with heliocentric distance, by contrast, is a robust feature that persists across both missions despite their different absolute normalizations, and the cross-comparison with independent observational and theoretical estimates supports a largely instrument-independent trend of correlation-scale growth with distance from the Sun.

The spectral steepening is the other central result of this work: the close agreement between the observed $-0.5$ steepening and previous FORWARD-modeling MHD simulations \citep{pecora2024mhdPUNCH} shows that turbulence parameters can be extracted from remote-sensing observations through a reproducible, predictable framework. This consistency opens the door to strong cross-mission synergies, offering a reliable route for mapping 2D projected brightness observations back onto the underlying 3D magnetohydrodynamic turbulence properties of the expanding solar wind and remote regions of the observable universe.

\section*{Acknowledgements}
PUNCH is a heliophysics mission to study the corona, solar wind, and space weather as an integrated system, and is part of NASA’s Explorers program (Contract 80GSFC14C0014).
This work was supported at the University of Delaware in part by the PUNCH project through subcontract N99054DS from NASA/SWRI to the University of Delaware.
This work was also partially supported by the U.S. National Science Foundation, award PHY-2108834, through the NSF/DOE Partnership in Basic Plasma Science and Engineering, by NSF SHINE grant PHYS312218 at the University of Delaware, and the Plan for NASA EPSCoR Research Infrastructure Development (RID) in Delaware (NASA award 80NSSC22M0039).
SEG acknowledges the National Center for Atmospheric Research, a major facility sponsored by the National Science Foundation under Cooperative Agreement No. 1852977.
This work is partially supported by Agenzia Spaziale Italiana (ASI) in the framework of the project ``Radial Evolution of large- and KInetic-scale Processes in the Expanding Solar wind (REKIPES)'' C83C25000900005.

\section*{Data Availability}

All PUNCH data are fully open to anyone for use with no restrictions at
\url{https://punch.space.swri.edu/punch_science_getdata.php}.



\bibliographystyle{mnras}

\begin{thebibliography}{}
\makeatletter
\relax
\def\mn@urlcharsother{\let\do\@makeother \do\$\do\&\do\#\do\^\do\_\do\%\do\~}
\def\mn@doi{\begingroup\mn@urlcharsother \@ifnextchar [ {\mn@doi@} {\mn@doi@[]}}
\def\mn@doi@[#1]#2{\def\@tempa{#1}\ifx\@tempa\@empty \href {http://dx.doi.org/#2} {doi:#2}\else \href {http://dx.doi.org/#2} {#1}\fi \endgroup}
\def\mn@eprint#1#2{\mn@eprint@#1:#2::\@nil}
\def\mn@eprint@arXiv#1{\href {http://arxiv.org/abs/#1} {{\tt arXiv:#1}}}
\def\mn@eprint@dblp#1{\href {http://dblp.uni-trier.de/rec/bibtex/#1.xml} {dblp:#1}}
\def\mn@eprint@#1:#2:#3:#4\@nil{\def\@tempa {#1}\def\@tempb {#2}\def\@tempc {#3}\ifx \@tempc \@empty \let \@tempc \@tempb \let \@tempb \@tempa \fi \ifx \@tempb \@empty \def\@tempb {arXiv}\fi \@ifundefined {mn@eprint@\@tempb}{\@tempb:\@tempc}{\expandafter \expandafter \csname mn@eprint@\@tempb\endcsname \expandafter{\@tempc}}}

\bibitem[\protect\citeauthoryear{Bailey, Bandyopadhyay, Habbal  \& Druckmüller}{Bailey et~al.}{2025}]{bailey2025measurement}
Bailey Z.,  Bandyopadhyay R.,  Habbal S.,   Druckmüller M.,  2025, \mn@doi [The Astrophysical Journal Letters] {10.3847/2041-8213/ada363}, 980, L20

\bibitem[\protect\citeauthoryear{Batchelor}{Batchelor}{1953}]{batchelor1953theory}
Batchelor G.~K.,  1953, The theory of homogeneous turbulence.
Cambridge university press

\bibitem[\protect\citeauthoryear{Bishop, Oughton, Parashar  \& Perrott}{Bishop et~al.}{2026}]{bishop2026direct}
Bishop M.~A.,  Oughton S.,  Parashar T.~N.,   Perrott Y.~C.,  2026, \mn@doi [Physics of Fluids] {10.1063/5.0310561}, 38, 025107

\bibitem[\protect\citeauthoryear{Blackman \& Tukey}{Blackman \& Tukey}{1958a}]{blackman1958measurementI}
Blackman R.~B.,  Tukey J.~W.,  1958a, \mn@doi [Bell System Technical Journal] {https://doi.org/10.1002/j.1538-7305.1958.tb03874.x}, 37, 185

\bibitem[\protect\citeauthoryear{Blackman \& Tukey}{Blackman \& Tukey}{1958b}]{blackman1958measurementII}
Blackman R.~B.,  Tukey J.~W.,  1958b, \mn@doi [Bell System Technical Journal] {https://doi.org/10.1002/j.1538-7305.1958.tb01530.x}, 37, 485

\bibitem[\protect\citeauthoryear{Bruno \& Carbone}{Bruno \& Carbone}{2005}]{brunocarbone2005review}
Bruno R.,  Carbone V.,  2005, \mn@doi [Living Reviews in Solar Physics] {10.12942/lrsp-2005-4}, 2, 4

\bibitem[\protect\citeauthoryear{Burch, Moore, Torbert  \& Giles}{Burch et~al.}{2016}]{burch2016MMS}
Burch J.~L.,  Moore T.~E.,  Torbert R.~B.,   Giles B.~L.,  2016, \mn@doi [Space Science Reviews] {10.1007/s11214-015-0164-9}, 199, 5

\bibitem[\protect\citeauthoryear{Chasapis et~al.,}{Chasapis et~al.}{2017}]{chasapis2017high}
Chasapis A.,  et~al., 2017, \mn@doi [The Astrophysical Journal Letters] {10.3847/2041-8213/aa7ddd}, 844, L9

\bibitem[\protect\citeauthoryear{Chen et~al.,}{Chen et~al.}{2020}]{chen2020evolution}
Chen C. H.~K.,  et~al., 2020, \mn@doi [The Astrophysical Journal Supplement Series] {10.3847/1538-4365/ab60a3}, 246, 53

\bibitem[\protect\citeauthoryear{Chhiber, Usmanov, Matthaeus  \& Goldstein}{Chhiber et~al.}{2021}]{chhiber2021large}
Chhiber R.,  Usmanov A.~V.,  Matthaeus W.~H.,   Goldstein M.~L.,  2021, \mn@doi [The Astrophysical Journal] {10.3847/1538-4357/ac1ac7}, 923, 89

\bibitem[\protect\citeauthoryear{Chhiber, Wang, Usmanov  \& Matthaeus}{Chhiber et~al.}{2026}]{chhiber2026dynamicalage}
Chhiber R.,  Wang Y.,  Usmanov A.~V.,   Matthaeus W.~H.,  2026, \mn@doi [The Astrophysical Journal Letters] {10.3847/2041-8213/ae7a4a}, 1005, L10

\bibitem[\protect\citeauthoryear{Colaninno et~al.,}{Colaninno et~al.}{2025}]{colaninno2025punchNFI}
Colaninno R.,  et~al., 2025, \mn@doi [Solar Physics] {10.1007/s11207-025-02506-z}, 300, 104

\bibitem[\protect\citeauthoryear{Cranmer \& van Ballegooijen}{Cranmer \& van Ballegooijen}{2012}]{cranmer2012proton}
Cranmer S.~R.,  van Ballegooijen A.~A.,  2012, \mn@doi [The Astrophysical Journal] {10.1088/0004-637X/754/2/92}, 754, 92

\bibitem[\protect\citeauthoryear{Cuesta et~al.,}{Cuesta et~al.}{2022}]{cuesta2022isotropization}
Cuesta M.~E.,  et~al., 2022, \mn@doi [The Astrophysical Journal Letters] {10.3847/2041-8213/ac73fd}, 932, L11

\bibitem[\protect\citeauthoryear{DeForest, Howard  \& Tappin}{DeForest et~al.}{2013}]{deforest2013thomson}
DeForest C.~E.,  Howard T.~A.,   Tappin S.~J.,  2013, \mn@doi [The Astrophysical Journal] {10.1088/0004-637X/765/1/44}, 765, 44

\bibitem[\protect\citeauthoryear{{DeForest}, {Matthaeus}, {Viall}  \& {Cranmer}}{{DeForest} et~al.}{2016}]{deforest2016fading}
{DeForest} C.~E.,  {Matthaeus} W.~H.,  {Viall} N.~M.,   {Cranmer} S.~R.,  2016, \mn@doi [\apj] {10.3847/0004-637X/828/2/66}, 828, 66

\bibitem[\protect\citeauthoryear{{DeForest}, {Howard}, {Velli}, {Viall}  \& {Vourlidas}}{{DeForest} et~al.}{2018}]{deforest2018highly}
{DeForest} C.~E.,  {Howard} R.~A.,  {Velli} M.,  {Viall} N.,   {Vourlidas} A.,  2018, \mn@doi [\apj] {10.3847/1538-4357/aac8e3}, 862, 18

\bibitem[\protect\citeauthoryear{DeForest et~al.,}{DeForest et~al.}{2026}]{deforest2026PUNCH}
DeForest C.~E.,  et~al., 2026, \mn@doi [Solar Physics] {10.1007/s11207-026-02608-2}, 301, 16

\bibitem[\protect\citeauthoryear{Escoubet, Schmidt  \& Goldstein}{Escoubet et~al.}{1997}]{escoubet1997cluster}
Escoubet C.~P.,  Schmidt R.,   Goldstein M.~L.,  1997, \mn@doi [Space Science Reviews] {10.1023/A:1004923124586}, 79, 11

\bibitem[\protect\citeauthoryear{Freeland \& Handy}{Freeland \& Handy}{1998}]{freeland1998data}
Freeland S.~L.,  Handy B.~N.,  1998, \mn@doi [Solar Physics] {10.1023/A:1005038224881}, 182, 497

\bibitem[\protect\citeauthoryear{{Frisch}}{{Frisch}}{1995}]{frisch1995turbulence}
{Frisch} U.,  1995, {Turbulence: the legacy of AN Kolmogorov}.
Cambridge university press

\bibitem[\protect\citeauthoryear{Gazis \& Lazarus}{Gazis \& Lazarus}{1983}]{gazis1983radial}
Gazis P.~R.,  Lazarus A.~J.,  1983, in Solar Wind Five. JPL, \url {https://ntrs.nasa.gov/citations/19840005038}

\bibitem[\protect\citeauthoryear{Gibson et~al.,}{Gibson et~al.}{2016}]{gibson2016forward}
Gibson S.~E.,  et~al., 2016, \mn@doi [Frontiers in Astronomy and Space Sciences] {10.3389/fspas.2016.00008}, 3

\bibitem[\protect\citeauthoryear{Goodwill et~al.,}{Goodwill et~al.}{2026}]{goodwill2026averaging}
Goodwill J.,  et~al., 2026, Averaging Effects on the Solar Wind Alfvén Mach Number: Implications for Switchbacks and Alfvén Transition (\mn@eprint {arXiv} {2607.28773}), \url {https://arxiv.org/abs/2607.28773}

\bibitem[\protect\citeauthoryear{Hahn, Hofmeister, Koukras  \& Savin}{Hahn et~al.}{2025}]{hahn2025measurement}
Hahn M.,  Hofmeister S.~J.,  Koukras A.,   Savin D.~W.,  2025, \mn@doi [The Astrophysical Journal] {10.3847/1538-4357/ae017f}, 991, 97

\bibitem[\protect\citeauthoryear{Hellinger, Matteini, {\v{S}}tver{\'a}k, Tr{\'a}vn{\'\i}{\v{c}}ek  \& Marsch}{Hellinger et~al.}{2011}]{hellinger2011heating}
Hellinger P.,  Matteini L.,  {\v{S}}tver{\'a}k {\v{S}}.,  Tr{\'a}vn{\'\i}{\v{c}}ek P.~M.,   Marsch E.,  2011, \mn@doi [Journal of Geophysical Research: Space Physics] {https://doi.org/10.1029/2011JA016674}, 116

\bibitem[\protect\citeauthoryear{Hollweg}{Hollweg}{1986}]{hollweg1986transition}
Hollweg J.~V.,  1986, \mn@doi [Journal of Geophysical Research: Space Physics] {https://doi.org/10.1029/JA091iA04p04111}, 91, 4111

\bibitem[\protect\citeauthoryear{Howard \& DeForest}{Howard \& DeForest}{2012}]{howard2012thomson}
Howard T.~A.,  DeForest C.~E.,  2012, \mn@doi [The Astrophysical Journal] {10.1088/0004-637X/752/2/130}, 752, 130

\bibitem[\protect\citeauthoryear{Howard et~al.,}{Howard et~al.}{2008}]{howard2008secchi}
Howard R.~A.,  et~al., 2008, \mn@doi [Space Science Reviews] {10.1007/s11214-008-9341-4}, 136, 67

\bibitem[\protect\citeauthoryear{Huang et~al.,}{Huang et~al.}{2023}]{huang2023oneoverfPSP}
Huang Z.,  et~al., 2023, \mn@doi [The Astrophysical Journal Letters] {10.3847/2041-8213/acd7f2}, 950, L8

\bibitem[\protect\citeauthoryear{Isaacs, Tessein  \& Matthaeus}{Isaacs et~al.}{2015}]{isaacs2015systematic}
Isaacs J.~J.,  Tessein J.~A.,   Matthaeus W.~H.,  2015, \mn@doi [Journal of Geophysical Research: Space Physics] {https://doi.org/10.1002/2014JA020661}, 120, 868

\bibitem[\protect\citeauthoryear{Kaiser, Kucera, Davila, St.~Cyr, Guhathakurta  \& Christian}{Kaiser et~al.}{2008}]{kaiser2008STEREO}
Kaiser M.~L.,  Kucera T.~A.,  Davila J.~M.,  St.~Cyr O.~C.,  Guhathakurta M.,   Christian E.,  2008, \mn@doi [Space Science Reviews] {10.1007/s11214-007-9277-0}, 136, 5

\bibitem[\protect\citeauthoryear{Kiyani, Osman  \& Chapman}{Kiyani et~al.}{2015}]{kiyani2015dissipation}
Kiyani K.~H.,  Osman K.~T.,   Chapman S.~C.,  2015, \mn@doi [Philosophical Transactions of the Royal Society A: Mathematical, Physical and Engineering Sciences] {10.1098/rsta.2014.0155}, 373, 20140155

\bibitem[\protect\citeauthoryear{Laurent et~al.,}{Laurent et~al.}{2026}]{laurent2026wfipunch}
Laurent G.~T.,  et~al., 2026, \mn@doi [Solar Physics] {10.1007/s11207-026-02706-1}, 301, 112

\bibitem[\protect\citeauthoryear{Li, Jiang, Wang, Verscharen, Zeng, Russell, Giles  \& Burch}{Li et~al.}{2020}]{li2020evolution}
Li H.,  Jiang W.,  Wang C.,  Verscharen D.,  Zeng C.,  Russell C.~T.,  Giles B.,   Burch J.~L.,  2020, \mn@doi [The Astrophysical Journal Letters] {10.3847/2041-8213/aba531}, 898, L43

\bibitem[\protect\citeauthoryear{Machlup}{Machlup}{1981}]{machlup1981earthquakes}
Machlup S.,  1981, Noise in Physical Systems, 614, 157

\bibitem[\protect\citeauthoryear{Matthaeus \& Goldstein}{Matthaeus \& Goldstein}{1986}]{matthaeus1986low}
Matthaeus W.~H.,  Goldstein M.~L.,  1986, \mn@doi [Phys. Rev. Lett.] {10.1103/PhysRevLett.57.495}, 57, 495

\bibitem[\protect\citeauthoryear{Matthaeus \& Velli}{Matthaeus \& Velli}{2011}]{matthaeus2011who}
Matthaeus W.~H.,  Velli M.,  2011, \mn@doi [Space Science Reviews] {10.1007/s11214-011-9793-9}, 160, 145

\bibitem[\protect\citeauthoryear{Matthaeus, Smith  \& Oughton}{Matthaeus et~al.}{1998}]{matthaeus1998dynamical}
Matthaeus W.~H.,  Smith C.~W.,   Oughton S.,  1998, \mn@doi [Journal of Geophysical Research: Space Physics] {https://doi.org/10.1029/97JA03729}, 103, 6495

\bibitem[\protect\citeauthoryear{Matthaeus, Smith  \& Bieber}{Matthaeus et~al.}{1999}]{matthaeus1999correlation}
Matthaeus W.~H.,  Smith C.~W.,   Bieber J.~W.,  1999, \mn@doi [AIP Conference Proceedings] {10.1063/1.58686}, 471, 511

\bibitem[\protect\citeauthoryear{Matthaeus, Weygand  \& Dasso}{Matthaeus et~al.}{2016}]{matthaeus2016ensemble}
Matthaeus W.~H.,  Weygand J.~M.,   Dasso S.,  2016, \mn@doi [Phys. Rev. Lett.] {10.1103/PhysRevLett.116.245101}, 116, 245101

\bibitem[\protect\citeauthoryear{McComas, Elliott, Schwadron, Gosling, Skoug  \& Goldstein}{McComas et~al.}{2003}]{mccomas2003three_Ulysses}
McComas D.~J.,  Elliott H.~A.,  Schwadron N.~A.,  Gosling J.~T.,  Skoug R.~M.,   Goldstein B.~E.,  2003, \mn@doi [Geophysical Research Letters] {https://doi.org/10.1029/2003GL017136}, 30

\bibitem[\protect\citeauthoryear{Monin \& Yaglom}{Monin \& Yaglom}{1971}]{MoninYaglomI}
Monin A.~S.,  Yaglom A.~M.,  1971, Statistical Fluid Mechanics, Vol. 1.
MIT Press, Cambridge, Mass.

\bibitem[\protect\citeauthoryear{Monin \& Yaglom}{Monin \& Yaglom}{1975}]{MoninYaglomII}
Monin A.~S.,  Yaglom A.~M.,  1975, Statistical Fluid Mechanics, Vol. 2.
MIT Press, Cambridge, Mass.

\bibitem[\protect\citeauthoryear{Montroll \& Shlesinger}{Montroll \& Shlesinger}{1982}]{montroll1982onef}
Montroll E.~W.,  Shlesinger M.~F.,  1982, \mn@doi [Proceedings of the National Academy of Sciences] {10.1073/pnas.79.10.3380}, 79, 3380

\bibitem[\protect\citeauthoryear{Pecora, Yang, Gibson, Viall, Chhiber, DeForest  \& Matthaeus}{Pecora et~al.}{2024}]{pecora2024mhdPUNCH}
Pecora F.,  Yang Y.,  Gibson S.,  Viall N.~M.,  Chhiber R.,  DeForest C.,   Matthaeus W.~H.,  2024, \mn@doi [Solar Physics] {10.1007/s11207-024-02382-z}, 299, 137

\bibitem[\protect\citeauthoryear{Pecora, Matthaeus, Greco, Dmitruk, Yang, Carbone  \& Servidio}{Pecora et~al.}{2025}]{pecora2025spacetime}
Pecora F.,  Matthaeus W.~H.,  Greco A.,  Dmitruk P.,  Yang Y.,  Carbone V.,   Servidio S.,  2025, \mn@doi [Proceedings of the National Academy of Sciences] {10.1073/pnas.2519635122}, 122, e2519635122

\bibitem[\protect\citeauthoryear{Pecora et~al.,}{Pecora et~al.}{2026}]{pecora2026general_ubmsh}
Pecora F.,  et~al., 2026, \mn@doi [Journal of Geophysical Research: Space Physics] {https://doi.org/10.1029/2025JA034919}, 131, e2025JA034919

\bibitem[\protect\citeauthoryear{Perez, Bourouaine  \& Dorseth}{Perez et~al.}{2026}]{perez2026integral}
Perez J.~C.,  Bourouaine S.,   Dorseth M.,  2026, \mn@doi [The Astrophysical Journal Letters] {10.3847/2041-8213/ae77f4}, 1005, L4

\bibitem[\protect\citeauthoryear{Perrone, Stansby, Horbury  \& Matteini}{Perrone et~al.}{2019}]{perrone2019radial}
Perrone D.,  Stansby D.,  Horbury T.~S.,   Matteini L.,  2019, \mn@doi [Monthly Notices of the Royal Astronomical Society] {10.1093/mnras/sty3348}, 483, 3730

\bibitem[\protect\citeauthoryear{Pope}{Pope}{2000}]{pope2000turbulent}
Pope S.~B.,  2000, Turbulent flows.
Cambridge university press

\bibitem[\protect\citeauthoryear{Pradata, Roy, Matthaeus, Wang, Chhiber, Pecora  \& Yang}{Pradata et~al.}{2025}]{pradata2025observations_apjl_mercury1f}
Pradata R.~A.,  Roy S.,  Matthaeus W.~H.,  Wang J.,  Chhiber R.,  Pecora F.,   Yang Y.,  2025, \mn@doi [The Astrophysical Journal Letters] {10.3847/2041-8213/adc9b2}, 984, L23

\bibitem[\protect\citeauthoryear{Prete, Perri, Meringolo, Primavera  \& Servidio}{Prete et~al.}{2025}]{prete2025interaction}
Prete G.,  Perri S.,  Meringolo C.,  Primavera L.,   Servidio S.,  2025, \mn@doi [The Astrophysical Journal Supplement Series] {10.3847/1538-4365/adb62f}, 277, 44

\bibitem[\protect\citeauthoryear{Pugachev}{Pugachev}{1965}]{pugachev1965theory}
Pugachev V.~S.,  1965, Theory of random functions and its application to control problems.
Oxford, New York

\bibitem[\protect\citeauthoryear{Roberts, Klein, Goldstein  \& Matthaeus}{Roberts et~al.}{1987}]{roberts1987nature}
Roberts D.~A.,  Klein L.~W.,  Goldstein M.~L.,   Matthaeus W.~H.,  1987, \mn@doi [Journal of Geophysical Research: Space Physics] {https://doi.org/10.1029/JA092iA10p11021}, 92, 11021

\bibitem[\protect\citeauthoryear{Ruiz, Dasso, Matthaeus  \& Weygand}{Ruiz et~al.}{2014}]{ruiz2014characterization}
Ruiz M.~E.,  Dasso S.,  Matthaeus W.~H.,   Weygand J.~M.,  2014, \mn@doi [Solar Physics] {10.1007/s11207-014-0531-9}, 289, 3917

\bibitem[\protect\citeauthoryear{Sharma \& Morton}{Sharma \& Morton}{2023}]{sharma2023transverse}
Sharma R.,  Morton R.~J.,  2023, \mn@doi [Nature Astronomy] {10.1038/s41550-023-02070-1}, 7, 1301

\bibitem[\protect\citeauthoryear{Taylor}{Taylor}{1938}]{taylor1938spectrum_frozenin}
Taylor G.~I.,  1938, \mn@doi [Proceedings of the Royal Society of London. Series A, Mathematical and Physical Sciences] {10.1098/rspa.1938.0032}, 164, 476

\bibitem[\protect\citeauthoryear{Telloni et~al.,}{Telloni et~al.}{2021}]{telloni2021evolution}
Telloni D.,  et~al., 2021, \mn@doi [The Astrophysical Journal Letters] {10.3847/2041-8213/abf7d1}, 912, L21

\bibitem[\protect\citeauthoryear{Telloni et~al.,}{Telloni et~al.}{2024}]{telloni2024metis}
Telloni D.,  et~al., 2024, \mn@doi [The Astrophysical Journal Letters] {10.3847/2041-8213/ad5a8c}, 973, L48

\bibitem[\protect\citeauthoryear{Vourlidas et~al.,}{Vourlidas et~al.}{2016}]{vourlidas2016WISPR}
Vourlidas A.,  et~al., 2016, \mn@doi [Space Science Reviews] {10.1007/s11214-014-0114-y}, 204, 83

\bibitem[\protect\citeauthoryear{Wang, Matthaeus, Chhiber, Roy, Pradata, Pecora  \& Yang}{Wang et~al.}{2024}]{wang2024noisePUNCH}
Wang J.,  Matthaeus W.~H.,  Chhiber R.,  Roy S.,  Pradata R.~A.,  Pecora F.,   Yang Y.,  2024, \mn@doi [Solar Physics] {10.1007/s11207-024-02401-z}, 299, 169

\bibitem[\protect\citeauthoryear{Wang, Pecora, Chhiber, Roy  \& Matthaeus}{Wang et~al.}{2026a}]{wang2025interplanetary_PNAS}
Wang J.,  Pecora F.,  Chhiber R.,  Roy S.,   Matthaeus W.~H.,  2026a, \mn@doi [Proceedings of the National Academy of Sciences] {10.1073/pnas.2519811122}, 123, e2519811122

\bibitem[\protect\citeauthoryear{Wang, Pecora, Chhiber, Pradata, Adhikari  \& Matthaeus}{Wang et~al.}{2026b}]{wang2026fnoise_superposition}
Wang J.,  Pecora F.,  Chhiber R.,  Pradata R.~A.,  Adhikari S.,   Matthaeus W.~H.,  2026b, \mn@doi [Monthly Notices of the Royal Astronomical Society] {10.1093/mnras/stag722}, 548, stag722

\bibitem[\protect\citeauthoryear{Wicks, Chapman  \& Dendy}{Wicks et~al.}{2008}]{wicks2008spatial}
Wicks R.~T.,  Chapman S.~C.,   Dendy R.~O.,  2008, \mn@doi [The Astrophysical Journal] {10.1088/0004-637X/690/1/734}, 690, 734

\bibitem[\protect\citeauthoryear{Wicks, Owens  \& Horbury}{Wicks et~al.}{2010}]{wicks2010variation}
Wicks R.~T.,  Owens M.~J.,   Horbury T.~S.,  2010, \mn@doi [Solar Physics] {10.1007/s11207-010-9509-4}, 262, 191

\bibitem[\protect\citeauthoryear{Wiener}{Wiener}{1930}]{wiener1930generalized}
Wiener N.,  1930, \mn@doi [Acta Mathematica] {10.1007/BF02546511}, 55, 117

\bibitem[\protect\citeauthoryear{Yordanova, Vaivads, Andr\'e, Buchert  \& V\"or\"os}{Yordanova et~al.}{2008}]{yordanova2008magnetosheath}
Yordanova E.,  Vaivads A.,  Andr\'e M.,  Buchert S.~C.,   V\"or\"os Z.,  2008, \mn@doi [Phys. Rev. Lett.] {10.1103/PhysRevLett.100.205003}, 100, 205003

\bibitem[\protect\citeauthoryear{Zhao, Zank, Telloni, Stevens, Kasper  \& Bale}{Zhao et~al.}{2022}]{zhao2022turbulent}
Zhao L.-L.,  Zank G.~P.,  Telloni D.,  Stevens M.,  Kasper J.~C.,   Bale S.~D.,  2022, \mn@doi [The Astrophysical Journal Letters] {10.3847/2041-8213/ac5fb0}, 928, L15

\makeatother
\end{thebibliography}

\end{document}